\documentclass[12pt]{article}
\usepackage{amsfonts}
\usepackage{amsmath}
\usepackage{epsfig}

\begin{document}

\begin{flushright}
08/2014\\
\end{flushright}
\vspace{5mm}
\begin{center}
\large {\bf A Microscopic Approach to Quark and Lepton Masses and Mixings}\\
\mbox{ }\\
\normalsize
\vspace{2.0cm}
{\bf Bodo Lampe} \\              
\vspace{0.3cm}
II. Institut f\"ur theoretische Physik der Universit\"at Hamburg \\
Luruper Chaussee 149, 22761 Hamburg, Germany \\
\vspace{3.0cm}
{\bf Abstract}\\
\end{center} 
In recent papers a microscopic model for the SM Higgs mechanism has been proposed, and an idea how to determine the 24 quark and lepton masses of all 3 generations has emerged in that framework. This idea is worked out in detail here by accommodating the fermion masses and mixings to microscopic parameters. The top quark mass can be given in terms of the Fermi scale and of certain exchange couplings of isospin vectors obeying a tetrahedral symmetry. 
The observed hierarchy in the family spectrum is attributed to a natural hierarchy in the microscopic couplings. The neutrinos will be shown to vibrate within the potential valleys of the system, thus retaining very tiny masses. This is related to a Goldstone effect inside the internal dynamics. A discussion of the quark and lepton mixing matrices is also included. The mixing angles of the PMNS matrix are calculated for an example set of parameters, and a value for the CP violating phase is given.



\newpage

\normalsize



\section{Introduction}





The Standard Model of elementary particles (SM) is very successful on the phenomenological level. The outcome of (almost) any particle physics experiment can be predicted accurately within this model, and, where not, by some straightforward extension. For example, one may introduce right handed neutrinos to account for tiny neutrino masses\cite{nuemix}.

Nevertheless, it is widely believed that the SM is only an effective low-energy theory valid below a certain energy scale $\Lambda_r$, which is supposed to be of the order of 1-10 TeV. This view is based on the fact that the SM has many unknown parameters and one rather mysterious component, the so-called Higgs field, which is needed for the spontaneous symmetry breaking (SSB) taking place in the model. Within the Higgs sector the most challenging part is the set of Yukawa couplings to fermions, which comprises the majority of the unknown parameters of the SM Lagrangian.

In recent papers a microscopic model of the Higgs mechanism has been developed\cite{laa,lamm,lamm1}, and also an idea how the quark and lepton states arise in that model. This concept will be used in the present article in an attempt to determine the fermion masses and mixings. 

At first sight, the spectrum of quarks and leptons seems difficult to explain, because it extends over many orders of magnitude, starting from the neutrinos with their tiny masses below 1 eV, passing over to the 'everyday life' particles e, u and d with masses of order $10^6$\ eV,  proceeding to muon and strange-quark (about $10^8$eV), ascending to charm, $\tau$ and bottom, which have masses of order $10^9$eV, and finishing with the top quark, whose mass of 1.7$\times 10^{11}$eV lies suspiciously close to the SSB scale $\Lambda_F$, the value of the Higgs mass and to twice the W-mass. For the neutrinos it is reasonable to believe that their mass might be some kind of higher order effect\cite{radi}, is protected by symmetry or generated by a variant of the popular seesaw mechanism\cite{seesaw,seevalle}. Other approaches to explain the hierarchy in the particle masses consider textures\cite{ramond} like 'democratic' mass matrices\cite{democ} with identical entries. These show the desirable feature that after diagonalization there is one very heavy particle (the top quark), and the rest have small masses. 

Unfortunately, a physical understanding of the underlying dynamics responsible for these effects is still lacking. For example, in (supersymmetric) grand unified theories fermion masses essentially remain free parameters. Furthermore, those models usually introduce many more additional degrees of freedom without much ambition to determine them from first principles. The point is that theories of that kind only extrapolate and extend the symmetries observed at low energies to small distances and that there is a strong amount of arbitrariness in this procedure. In my opinion it is obvious that a physical understanding of the masses and mixings is only possible in a microscopic theory. Superstring theories seem to offer such an understanding. However, although 'in principle' able to determine the masses as energies of string excitations, to my knowledge they have not come up with definite and verifiable predictions.

The present study is devoted to partially fill this gap. Some of the above mentioned 'textures' will reappear in the sections below. For example, neutrinos are indeed protected by the symmetries of new interactions. Furthermore, a kind of democratic texture will be derived which makes the top quark the heaviest fermion. More precisely, it arises from a symmetry breaking contribution modified by a Dzyaloshinskii-Moriya component\cite{dmgut} in such a way that all entries of the mass matrix effectively get identical  contributions. The appearance of this modification turns out to be a reflection of the $SU(2)_L$ gauge symmetry (breaking) on the microscopic level.

A related question is how the mixing between the generations can be understood. Most of the mixing angles are now known with a reasonable accuracy\cite{ckm,nuemix}. In particular, there is a hierarchy in the mixing matrix for the quark sector, but not in the lepton sector. Approaching the mixing problem in the present model I will be able to give some preliminary results mainly for the neutrinos and also set up the environment to derive the CKM mixing angles. In the course of the discussion an explanation will be given, of why the quark mixings are naturally 'small' and the lepton mixings naturally 'large'.


\section{Quarks and Leptons as Isospin Excitations of a Tetrahedral Shubnikov Group}

The Higgs doublet of the Standard Model can be parametrized as
\begin{eqnarray}
\Phi=\frac{1}{\sqrt{2}}
\begin{pmatrix}
i(\pi_x-i \pi_y) \\
\sigma -i\pi_z
\end{pmatrix}
\label{hig77}
\end{eqnarray}
where $\sigma = \Lambda_F +\phi$ acquires a vacuum expectation value $\langle \sigma \rangle =\Lambda_F=\sqrt{\frac{\mu^2}{\lambda}}=246$GeV through the form of the potential
\begin{eqnarray}  
V(\Phi)&=&-\mu^2 \Phi^+ \Phi + \lambda (\Phi^+\Phi)^2= -\frac{1}{2}\mu^2 (\sigma^2 +\vec\pi^2)+\frac{1}{4}\lambda (\sigma^2 +\vec\pi^2)^2 \nonumber \\
&=&\frac{1}{4}\lambda [-\Lambda_F^4+4\Lambda_F\phi\vec\pi^2+4\Lambda_F^2\phi^2+4\Lambda_F\phi^3+\phi^4+\vec\pi^4+2\phi^2\vec\pi^2]
\label{hipo}
\end{eqnarray}
and $\vec \pi =(\pi_x,\pi_y,\pi_z)$ gets 'eaten' by the longitudinal modes of the afterwards massive W-bosons. This can be made explicit by a SU(2) gauge transformation of the form\cite{technireview} 
\begin{eqnarray}  
U=\exp (\frac{i\vec \tau \vec \pi}{2\Lambda_F}  )
\label{hipoga}
\end{eqnarray}
which formally removes $\vec \pi$ from the Higgs doublet.

\begin{figure}
\begin{center}
\epsfig{file=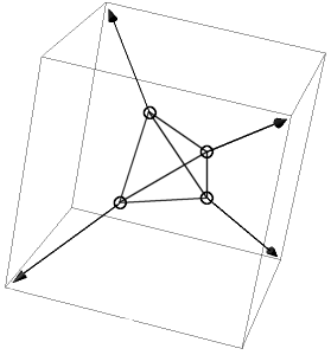,height=6.0cm}
\caption{The local ground state of the model, living in a 3-dimensional internal $R^3$ space. Shown are the corner points (small circles) of the internal tetrahedron, which can be represented by their coordinate vectors $\vec r_i$. The origin of coordinates is taken to be the center of the tetrahedron, and is identical to the base point of the fiber in Minkowski space. 
On each corner point $i=1,2,3,4$ there is an axial spin vector $\vec \pi_i$, pointing in the same radial direction as $\vec r_i$.}
\nonumber
\end{center}
\end{figure}

The present calculation is motivated by a Nambu-Jona-Lasinio (NJL) type of interpretation of the SSB mechanism. This heuristic ansatz will then be vigorously extended to a new, microscopic picture of the SM particle system. The starting point is an isospin pair $\psi=(U,D)$ of Dirac fermions distantly similar as in technicolor models\cite{techni1,techni2,techni3,techni4}, however without a technicolor quantum number. Rather we shall assume that the pairing mechanism is due to exchange interactions and strong correlations between fermions, effects which in many body physics are known to be responsible for SSB in superconductors and (anti)-ferromagnets. In contrast to solid state physics we do not consider these effects in physical space, but attribute them to arise from an independent dynamics which is active in the internal spaces. It is this dynamics which will allow us to put hand on the values of the fermion masses.

The main idea is that (weak) isospin arises from a nonrelativistic internal $R^3$ space much like ordinary spin arises from physical space. In other words, the internal space is assumed to possess a rotational SO(3)-symmetry for which the doublet $\psi=(U,D)$ serves as an (internal) Pauli spinor with an initial internal SU(2) spin symmetry. These internal spins are assumed to undergo interactions in the internal space which can be described by internal Heisenberg spin interactions which are formally similar to those describing spin interactions in solids. 

The geometrical picture is that the world is a fiber bundle over Minkowski space with fibers given by internal $R^3$ spaces, and that within these fibers physical processes take place, which can be described by a higher dimensional quantum electrodynamics. 
This idea was worked out in ref. \cite{laa} and the interested reader is referred to that paper for more details. It is further assumed that at high temperatures there is a symmetric phase in which the internal spins are distributed randomly in the fibers, giving rise to a local internal SU(2) symmetry of the Lagrangian, local in the sense that on each site in each fiber the spins may be rotated independently. 

When the temperature of the universe decreases from big bang energies to TeV values the internal magnetic interactions within the fiber lead to the frustrated\cite{frust} tetrahedral structure shown in fig. 1, and when it falls below the Fermi scale all the tetrahedrons over Minkowski space align as in fig. 2, a process which in ref.\cite{laa} was claimed to be the microscopic origin of the Higgs mechanism. A pairing process for the formation of the Higgs particle has also been described in that paper.

The configuration fig. 1 is the starting point of the present calculation, because it is considered as the {\it local} ground state of the system. In other words, it is assumed that in each of the 3-dimensional internal $R^3$ fibers there is a discrete tetrahedral structure and that the internal dynamics is such that spin vectors arrange themselves according to this internal tetrahedral configuration. The tetrahedron itself has the tetrahedral group $S_4$ as point group symmetry. However, due to the pseudovector property of the internal spin vectors the whole system loses its reflection symmetries and obtains instead the Shubnikov symmetry group $A_4 + S ( S_4 - A_4)$\cite{shub,white,borov}, where S is the internal time reversal operation and $A_4$ is the subgroup of $S_4$ which does not contain reflections. Note that S itself does not belong to the Shubnikov group, and also the internal reflections do not. The Shubnikov group is chiral, the configuration with opposite chirality being given when the 4 spin vectors would point inwards instead of outwards. Before the formation of the chiral tetrahedron the internal spins U and D, which according to eq. (\ref{njl09y}) are the building blocks of the spin vectors $\vec \pi_i$, can freely rotate and thus there is an internal spin SU(2) symmetry group, which however is broken to $A_4 + S ( S_4 - A_4)$ when the chiral tetrahedron is formed.

With respect to (external) Lorentz symmetry both U and D can appear as lefthanded or righthanded objects, so that one may in fact consider separately a $SU(2)_L$ for the lefthanded and $SU(2)_R$ for the righthanded objects. Before the advent of the chiral tetrahedrons and of the gauge bosons the Higgs sector of the SM is symmetric under $SU(2)_L \times SU(2)_R$, and a Nambu-Jona-Lasinio (NJL) model with this symmetry may we formulated.

\begin{figure}
\begin{center}
\epsfig{file=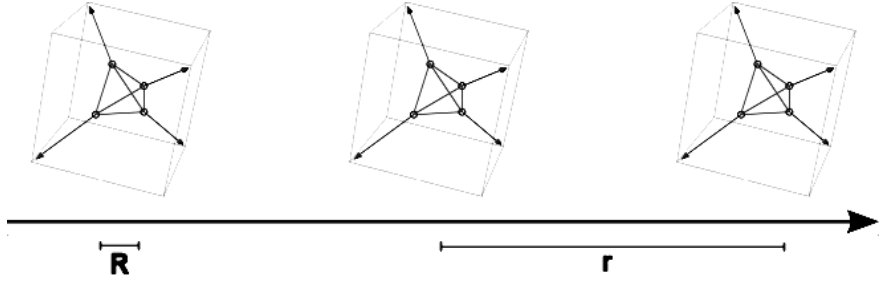,height=4.4cm}
\bigskip
\caption{The global ground state of the model after SSB consists of an aligned system of chiral tetrahedrons over Minkowski space (the latter represented by the long arrow). R is the magnitude of one tetrahedron and r the distance between two of them. Note that contrary to what is drawn here, the tetrahedra extend into internal space alone, not into Minkowski space. Before the SSB the chiral tetrahedrons are oriented randomly (not shown) and there is a corresponding {\it local}  $SO(3)$ symmetry, because each rigid tetrahedron can be rotated freely and independently from the others.}
\nonumber
\end{center}
\end{figure}

The NJL philosophy operates as follows: the SSB is induced by formation of bound states and condensates of the fermion doublet $\psi=(U,D)$. Namely, the quadratic part 
\begin{eqnarray}  
V_2(\Phi)=-\mu^2 \Phi^+ \Phi= -\frac{1}{2}\mu^2 (\sigma^2 +\vec\pi^2)
\label{hipo700}
\end{eqnarray}
of the potential eq. (\ref{hipo}) is equivalent to a NJL interaction of the form 
\begin{eqnarray}
V_{NJL}=G_{NJL}[ (\bar \psi \psi)^2+(\bar \psi i \gamma_5 \vec \tau \psi)^2] 
\label{njl09}
\end{eqnarray}
where $G_{NJL}$ denotes the NJL coupling strength, which in the SSB regime, where $V_2(\Phi)<0$, must be negative as well. (Note that $\bar \psi i \gamma_5 \vec \tau \psi$ is always real.) In fact using the method of auxiliary fields one can show $V_2(\Phi)=V_{NJL}$ provided one chooses 
\begin{eqnarray}
\sigma &=&-2G_{NJL}\bar \psi \psi \nonumber \\
\vec\pi &=&-2G_{NJL}\bar \psi i \gamma_5 \vec \tau \psi
\label{njl09y}
\end{eqnarray}
and thus obtain a sigma model from the original NJL potential (\ref{njl09}).

For later use I will rewrite eq. (\ref{njl09}) as 
\begin{eqnarray}
V_{NJL}=-J_F [ (\frac{\bar \psi \psi}{\mu^3})^2+(\frac{\bar \psi i \gamma_5 \vec \tau \psi}{\mu^3})^2]  
\label{njl229}
\end{eqnarray}
so that in brackets there are dimensionless quantities and $J_F=-G_{NJL}\mu^6$ has the dimension 4 of an energy density. 

This equation has been interpreted in ref. \cite{laa} to describe the dynamics of interacting chiral internal spin vectors $\vec \pi$, normalized to the SSB scale $\mu$.
Here 'chiral' refers both to internal and physical space, because $\gamma_5$ is a building block for chiral objects in physical space, while the appearance of (internal) Pauli matrices $\vec \tau$ signals chiral objects in (internal) space. Moreover, in the framework of the internal Heisenberg spin theory $J_F$ is to be interpreted as the internal exchange energy density corresponding to certain exchange integrals over internal $R^3$ space to be described later. In the SSB regime one has $J_F>0$ (because of $G_{NJL}<0$) corresponding to a ferromagnetic interaction. This interaction accounts for neighbouring tetrahedrons aligning themselves over Minkowski space as shown in fig. 2. 

There is a slight complication on these considerations, because at SSB energies after redefinition of $\sigma = \Lambda_F +\phi$, the $\vec\pi$-$\vec\pi$ interaction eq. (\ref{hipo700}) seems to disappear, because the sum of terms $\sim \vec\pi^2$ vanishes in the potential $V(H)$ as made explicit by the last of eqs. (\ref{hipo}). However, when the $\vec \pi$ triplet is absorbed as the longitudinal mode of the $\vec W$-boson the internal Heisenberg spin interaction reappears as part of the mass term $m_W^2 W_\mu W^\mu$. The alignment of tetrahedrons in fig. 2 will then experience a modification which is dictated by the gauge symmetry of the fiber bundle formed by all tetrahedrons. As a consequence the ferromagnetic Heisenberg interaction has to be modified by a Dzyaloshinskii-Moriya component\cite{dmgut,dmz1,dmz2} in this regime. Details will be given in section 5. 

Next I want to extent the view to small distances and high energies. At high energies, there is no SSB and instead of the negative potential term $V_2$ one has a strictly positive potential, which still can be described by eq. (\ref{njl09}), however with a positive coupling $G_{NJL}$. Rewriting that equation as 
\begin{eqnarray}
V_{NJL}=-J_A [ (\frac{\bar \psi \psi}{\Lambda_r^3})^2+(\frac{\bar \psi i \gamma_5 \vec \tau \psi}{\Lambda_r^3})^2] 
\label{njl229a}
\end{eqnarray}
one should not take the SSB scale $\mu$ as normalization scale any more. Instead another reference scale $\Lambda_r \gg \mu$ has to be introduced which physically corresponds to the distance between two tetrahedrons (alternatively one could utilize the extension of one of them). 
One thus obtains an antiferromagnetic internal spin interaction with a negative exchange coupling $J_A$. This repulsion effect leads to the frustrated antiferromagnetic vacuum structure shown in fig. 1. $J_F$ and $J_A$ differ because they correspond to exchange integrals over different regions of space. $J_A$ is dominated by an integral over the volume of one internal tetrahedron (leading to the frustrated internal antiferromagnetic configuration), while $J_F$ is the exchange integral for spin vectors of different tetrahedrons (leading to the aligment of different tehtrahedrons over Minkowski space).

The situation is reminiscent to the theory of ordinary magnets, where the exchange coupling integral J is known to vary with the distance. At large lattice spacings and corresponding large distances between spin vectors ( much larger than the extension of the electron wave function) one has $J>0$ and a ferromagnetic behavior. On the other hand, antiferromagnets like Cr and Mn are characterized by small lattice spacings and corresponding small distances between spin vectors, typically not much larger than the extension of the electron wave function. In these cases $J<0$, i.e. antiferromagnetic behavior.


According to fig. 1 there is one chiral internal spin vector $\vec \pi_i$ for each of the 4 constituents of the internal tetrahedron. In the ground state these vectors point radially away from the origin. Their sum
\begin{equation}
\langle\vec \pi\rangle = \sum_{i=1}^4 \langle\vec \pi_i\rangle
\label{mc7}
\end{equation}
vanishes corresponding to a vanishing vev $\langle\vec \pi \rangle=0$ in accordance with the SM vacuum structure of the Higgs doublet (\ref{hig77}). Excited states arise as vibrations of the vectors $\vec \pi_i$ in fig. 1 and will be interpreted as quarks and leptons. They can be classified according to the system's symmetry group, the Shubnikov group $A_4+S(S_4-A_4)$. This group, which remains unbroken at low energies, has only 1- and 3- dimensional representations, i.e. singlets (interpreted as leptons) and triplets (interpreted as the 3 colors of quarks).

\section{Chiral Extension of the Model}

When studying the dynamics of the internal spin vectors to derive the spectrum of the excited states one notes that $\vec \pi \sim \bar \psi i \gamma_5 \vec \tau \psi$ is not a quantity simple to handle, because it does not fulfill the canonical commutation relations for spin vectors. Secondly, it turns out that the internal Hamiltonian (related to the internal time variable) cannot be written in terms of $\vec \pi$. The appropriate internal vector to use is $\psi^\dagger \gamma_5 \vec \tau \psi$, a well known conserved charge observable from current algebra\cite{current}. However, due to the factor of $\gamma_5$ these vectors still do not fulfill the usual angular momentum commutation relations because their commutator is a scalar, and not a pseudo-scalar any more. In order that the algebra of internal spin vectors closes one is forced to consider the following linear combinations of internal spin vectors
\begin{eqnarray}
\vec Q_R= \frac{1}{\Lambda^3} \psi^\dagger (1+\gamma_5) \vec \tau \psi=a_R^\dagger \vec \tau a_R   - b_L^\dagger  \vec \tau b_L  \\
\vec Q_L=\frac{1}{\Lambda^3} \psi^\dagger (1-\gamma_5) \vec \tau \psi= a_L^\dagger  \vec \tau a_L   - b_R^\dagger  \vec \tau b_R  
\label{m1u1}
\end{eqnarray} 
where $a(^\dagger)_{L,R}$ and $b(^\dagger)_{L,R}$ denote doublets of annihilation(creation) operators of the fundamental fermion and anti-fermion with left and right handed polarization. 
For later use one should discriminate between particle and antiparticle components
\begin{eqnarray}
\vec S&=& a_L^\dagger  \vec \tau a_L \qquad \qquad 
\vec T= b_L^\dagger  \vec \tau b_L \label{njl4baa}\\
\vec U&=& a_R^\dagger \vec \tau a_R  \qquad \qquad 
\vec V= b_R^\dagger  \vec \tau b_R \label{njl4b87}
\end{eqnarray} 
These vectors fulfill the canonical algebra of SU(2) generators\cite{current}. The point to note here is the impact of the chiral symmetry group $SU(2)_R\times SU(2)_L$, because in current algebra $\vec Q_L$ and $\vec Q_R$ correspond to the conserved charges of the $SU(2)_R\times SU(2)_L$ symmetry, the factor $\Lambda^{-3}$ arising from the spatial integral of the time component of the left and right handed currents. $\vec S$ and $\vec V$ can be considered as 2 alternative sets of $SU(2)_L$ generators, while $\vec T$ and $\vec U$ serve the same purpose for $SU(2)_R$. For the question of the fermion masses addressed in the next sections one may restrict to consider the vectors $\vec S$ and $\vec T$ because this corresponds to $\bar \psi_R \psi_L$ of the fermion mass term, from which the $\bar \psi_L \psi_R$ part can be obtained by complex conjugation. 

Technically, restricting to $\vec S$ and $\vec T$ one is associating the transition between $SU(2)_L$ and $SU(2)_R$ to the charge conjugation operator\cite{nesti}. Alternatively, one could associate it to parity, by restricting to $\vec S$ and $\vec U$. However, this would not lead to the mass terms $\sim \bar \psi_R \psi_L$, and moreover the pairs of Shubnikov states introduced later in eq. (\ref{eq833hg}) could not be interpreted as partners of an $SU(2)_L$ doublet. As well known from the theory of SU(2) representations for a particle operator of the form $a=(U,D)$ the corresponding antiparticle operator is $b=(-D^c,U^c)$, i.e. transitions between particles and antiparticles involve a simultaneous interchange of isospin quantum numbers.

$\vec S$ and $\vec T$ transform under $SU(2)_L$ and $SU(2)_R$, respectively, and fulfill the commutation relations of two decoupled angular momenta in the sense that 
\begin{eqnarray}
[S_a,S_b]=i\epsilon_{abc}S_c \qquad 
[T_a,T_b]=i\epsilon_{abc}T_c \qquad
[S_a,T_b]=0 
\label{sv1}
\end{eqnarray}
On the other hand, within the internal dynamics advocated in this paper $\vec S$ and $\vec T$ play the role of angular momentum observables corresponding to rotations of the internal $R^3$ space. The fact that they are dimensionless according to eqs. (\ref{njl4baa}) agrees with this interpretation, because an angular momentum $\vec r \times \vec p$ is always dimensionless. Physically, the scale $\Lambda$ can be identified as $\Lambda=\mu$ in the SSB regime and $\Lambda_r$ at high energies. 

Using eq. (\ref{njl4baa}) one is effectively including further dynamical vibrators $\sim \psi^\dagger \vec \tau \psi$ in addition to the axial vectors $\vec \pi$. This kindly solves another problem not discussed so far, namely the 4$\times$3 d.o.f. of the internal spin vectors in fig. 1 yield only 12 excitation states instead of the necessary 24 quarks and leptons. In order to obtain the remaining 12 (which turn out to be their isospin partners), in ref. \cite{laa} it was proposed that internal displacive vibrations should be included in addition to spin wave excitations. In the present context the doubling of the number of excitations is obtained without displacive vibrations by going to the closed algebra eq. (\ref{mm4xx1}) of the 8 internal spin vectors $\vec S_i$ and $\vec T_i$, i=1-4, whose vacuum values are depicted in fig. 3. 

In that figure the vectors $\langle \vec S_i\rangle $ are shown pointing outwards and $\langle \vec T_i\rangle $ pointing inwards fulfilling
\begin{eqnarray} 
\langle \vec S_i\rangle =  -\langle \vec T_i\rangle   
\label{nba77}
\end{eqnarray}
If $\vec S_i$ and $\vec T_i$ were identical observables, this configuration would possess an internal time reversal symmetry (with symmetry group the 'grey' group $S_4 \times \{ 1,S\}$), because the time reversal invariance broken by the set of vectors pointing outwards would be restored by those pointing inwards. However, since $\vec S$ and $\vec T$ are physically different, the ground state has still the original Shubnikov group $A_4+S(S_4-A_4)$ as symmetry.

Eq. (\ref{nba77}) implies that the ground state gets contributions only from the $\gamma_5$ terms in eq. (\ref{njl4baa}) and $\langle \psi^\dagger \vec \tau \psi\rangle_i$ vanishes in the vacuum for each of the constituents $i=1,2,3,4$. In principle the opposite situation is conceivable as well, namely that $\langle \psi^\dagger \vec \tau \psi\rangle_i \neq 0$ while $\langle \psi^\dagger \gamma_5\vec \tau \psi\rangle_i$ vanishes. In that case one would have
\begin{eqnarray} 
\langle \vec S_i\rangle =  +\langle \vec T_i\rangle   
\label{nba77p}
\end{eqnarray}
a perfectly reasonable configuration, which maintains the Shubnikov symmetry for the system of 8 spin vectors as long as all internal vectors point e.g. outwards, in radial directions. In the numerical analysis presented in the following sections the configuration (\ref{nba77p}) will actually be preferred because technically it is easier to handle. In other words, eq. (\ref{nba77p}) will be used as equilibrium conditions for the concrete calculations of masses and eigenstates carried out in the next sections, cf. eq. (\ref{eqq23}).

The 24 eigenmodes of the system can be arranged in six 1-dimensional and six 3-dimensional representations of the Shubnikov group $A_4 + S ( S_4 - A_4)$\cite{shub,borov,white} to yield precisely the multiplet structure of the 24 quark and lepton states of the 3 generations, not less and not more. 
\begin{eqnarray} 
A_{\uparrow}(\nu_{e})+A_{\uparrow}(\nu_{\mu}) +A_{\uparrow}(\nu_{\tau}) &+& 
T_{\uparrow}(u)+T_{\uparrow}(c)+T_{\uparrow}(t)+ \nonumber \\
A_{\downarrow}(e)+A_{\downarrow}(\mu)+A_{\downarrow}(\tau) &+& 
T_{\downarrow}(d)+T_{\downarrow}(s)+T_{\downarrow}(b) 
\label{eq833hg}
\end{eqnarray}
Here $A_{\uparrow,\downarrow}$ and $T_{\uparrow,\downarrow}$ denote singlet and triplet representations of the Shubnikov group. For simplicity of notation, eq. (\ref{eq833hg}) does not distinguish between particle and anti-particle excitations. As shown later, the $\uparrow$ excitations can be obtained from the $\downarrow$ excitations by interchanging the roles of $\vec S$ and $\vec T$, which according to eq. (\ref{njl4baa}) involves an exchange of particle and anti-particle degrees of freedom. It is well known that for such an exchange the SU(2) quantum numbers have to be exchanged as well, because given an isospin doublet $(u,d)$ the corresponding doublet of anti-particles will be $(-\bar d,\bar u)$. Therefore, apart from describing anti-particles, the $\uparrow$ and $\downarrow$ excitations also describe isospin partners. This is an important feature which carries the isospin quantum number of the fundamental fermion $\psi$ to the quark and lepton isospin wave excitation.

\begin{figure}
\begin{center}
\epsfig{file=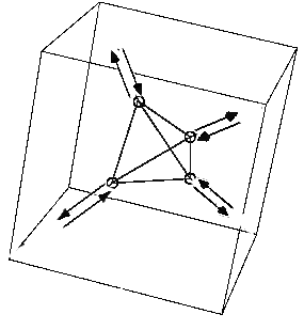,height=5.4cm}
\caption{The local ground state of the generalized NJL-model eqs.(\ref{n110c}) and (\ref{nba77}). The total of 8 internal spin vectors accounts for 3$\times$8 d.o.f. corresponding to 24 spin vibrations which can be classified according to the multiplet structure of the Shubnikov group, eq.(\ref{eq833hg}). The vectors $\vec S_i^0$ are assumed to point outwards and $\vec T_i^0=-\vec S_i^0$ inwards. According to eq. (\ref{njl4baa}) this corresponds to $\langle \psi^\dagger \gamma_5\vec \tau \psi \rangle_i\neq 0$. The alternative vacuum configuration (\ref{nba77p}) where $\langle \psi^\dagger \vec \tau \psi \rangle_i\neq 0$ and the $\vec T_i^0$ are parallel to the $\vec S_i^0$ instead of anti-parallel is not drawn.}
\nonumber
\end{center}
\end{figure}


In the framework of the chiral NJL dynamics the introduction of the second spin vector corresponds to introducing an additional term including $\bar\psi \vec \tau \psi$ into the potential. Actually it means to consider the most general $SU(2)_L \times SU(2)_R$ invariant potential based on a fundamental isospin doublet $\psi =(U,D)$, the general 2-flavor NJL model\cite{njl2,schechter} 
\begin{eqnarray}
V_{2NJL} &=& V_+ +V_- \nonumber \\
V_+&=&-J_+ [ (\frac{\bar \psi \psi}{\Lambda^3})^2+(\frac{\bar \psi \vec \tau \psi}{\Lambda^3})^2
+(\frac{\bar \psi i \gamma_5 \psi}{\Lambda^3})^2+(\frac{\bar \psi i \gamma_5 \vec \tau \psi}{\Lambda^3})^2] \nonumber \\
V_-&=&-J_-[ (\frac{\bar \psi \psi}{\Lambda^3})^2-(\frac{\bar \psi \vec \tau \psi}{\Lambda^3})^2
-(\frac{\bar \psi i \gamma_5 \psi}{\Lambda^3})^2+(\frac{\bar \psi i \gamma_5 \vec \tau \psi}{\Lambda^3})^2] 
\label{n110c}
\end{eqnarray}
$V_+$ and $V_-$ are separately chiral $SU(2)_L\times SU(2)_R$ invariant and in addition possess a $U(1)_V$ fermion number symmetry. Furthermore, $V_+$ is invariant under axial $U(1)_A$ transformations, while $V_-$ explicitly breaks this symmetry and can only be used if an axial anomaly is present (see below). The same scale $\Lambda$ ($=\mu$ oder $\Lambda_r$) as in eq. (\ref{njl4baa}) has been introduced to make the fermion operators dimensionless. As before, the NJL couplings can be written in terms of exchange energy densities $J_\pm$, and one needs $J_\pm >0(<0)$ to obtain  internal (anti)ferromagnetic behavior.

Rewriting $V_{2NJL}$ as
\begin{eqnarray}
V_{2NJL} =&-&(J_+ +J_-) [ (\frac{\bar \psi \psi}{\Lambda^3})^2
+(\frac{\bar \psi i \gamma_5 \vec \tau \psi}{\Lambda^3})^2] \nonumber \\
&-&(J_+ - J_-)[ (\frac{\bar \psi i \gamma_5 \psi}{\Lambda^3})^2 +
(\frac{\bar \psi \vec \tau \psi}{\Lambda^3})^2] 
\label{n110d}
\end{eqnarray}
and using the method of auxiliary fields one can transform the theory in the SSB region into a sigma-model, similar to eq. (\ref{njl09y}), by identifying
\begin{eqnarray}
\sigma &=&-2(G_+ +G_-)\bar \psi \psi \nonumber \\
\vec\pi &=&-2(G_+ +G_-)\bar \psi i \gamma_5 \vec \tau \psi \nonumber \\
\eta &=&-2(G_+ -G_-)\bar \psi i \gamma_5 \psi \nonumber \\
\vec v &=&-2(G_+ -G_-)\bar \psi \vec \tau \psi
\label{njl09w}
\end{eqnarray}
i.e. a scalar iso-scalar field (the physical Higgs $\sigma=\Lambda_F+\phi$), a pseudo-scalar iso-vector (the would be Goldstone bosons $\vec \pi$ absorbed by the weak bosons), a pseudo-scalar iso-scalar $\eta$ and a scalar iso-vector triplet $\vec v$ consisting of 2 charged fields $v^\pm$ and a neutral $v_{z}$. 

In other words, the requirement of a closed algebra of interacting spin vectors enforces the introduction of additional scalars which extend the Higgs sector of the SM to a specific two Higgs doublet model\cite{schechter}.
More precisely, the field content indicates a second scalar doublet formed by $\eta$ and $\vec v$ in addition to the ordinary Higgs doublet. However, in order to avoid a chiral ($\gamma_5$) vacuum structure of physical space, the $\eta$-field should not acquire a vacuum expectation value, 
so that the second doublet will not be part of the SSB-process, with negative mass terms, a non-trivial minimum of the potential and so on. 

Although technically possible, one should not put the octet of fields eq. (\ref{njl09w}) in the adjoint representation 
\begin{eqnarray}
\Sigma=\frac{1}{\sqrt{2}}
\begin{pmatrix}
\frac{1}{\sqrt{2}}(\sigma + v_z) & v^+ \\
v^- & \frac{1}{\sqrt{2}}(\sigma - v_z)
\end{pmatrix}
+\frac{i}{\sqrt{2}}
\begin{pmatrix}
\frac{1}{\sqrt{2}}(\eta + \pi_z) & \pi^+ \\
\pi^- & \frac{1}{\sqrt{2}}(\eta - \pi_z)
\end{pmatrix}
\label{higuu}
\end{eqnarray}
of a model with a larger $U(2)_R\times U(2)_L$ symmetry\cite{u2xu2}. The point is that the universal mass term tr$(\Sigma^\dagger \Sigma)$ of such a model would imply $J_-=0$ in eq. (\ref{n110d}). If any, this would be useful only as long as there are no axial anomalies in the theory which break the $U(1)_{A}$ subgroup of $U(2)_R\times U(2)_L$. 

As explained in ref \cite{laa} the underlying theory of the present model exhibits such an anomaly. 
In general this anomaly not only allows a non-vanishing value for $J_-$ but also makes the mass of the $\eta$ and $\vec v$ much larger than that of the weak gauge bosons and of the Higgs field\cite{u2xu2}. Those bound states may appear as heavy resonances in the TeV regime. They could be interesting dark matter candidates\cite{darkmatter,inert} and play a role at higher energies or higher temperatures of the universe. Their phenomenology, however, will not be discussed at this point, their d.o.f.s just being used as part of the components of the internal spin vectors whose vibrations give the quark and lepton mass spectrum.

\section{Masses and Mixings from Isospin Wave Equations}

The model set up in the last section will now be applied to calculate the quark and lepton masses. The idea is that masses can be identified with eigenfrequencies of excitations of the isospin vectors $\vec S$ and $\vec T$ eqs. (\ref{njl4baa}) and that these eigenfrequencies get contributions both from inner- and from inter-tetrahedral interactions. The {\it inner}-tetrahedral interactions are antiferromagnetic in nature and responsible for the frustrated tetrahedral configuration figs. 1 and 3, i.e. for the structure of the local vacuum. They are small distance contributions and relatively simple to treat because they can be described by an internal antiferromagnetic Heisenberg Hamiltonian for one tetrahedron alone, with corresponding internal spin vector excitations.

On the other hand there are {\it inter}-tetrahedral interactions fed by the 'ferromagnetic' SSB interactions between different tetrahedrons. Their leading effect turns out to be a contribution of order $O(\Lambda_F)$ solely to the top quark mass. Physically speaking, this interaction handicaps the specific eigenmode describing the top quark, because this mode disturbs the SSB alignment in the strongest possible way. Mathematically, the effect will be described by adding an effective universal term to the inner-tetrahedral Heisenberg interaction with a normal ferromagnetic plus a Dzyaloshinskii-Moriya component\cite{dmgut}. The sum of the 2 components will yield a quasi-democratic mass matrix which in leading order only contributes a term of order $\Lambda_F$ to the top-quark mass and nothing to the masses of the other quarks and leptons.

Let me start with the high energy / small distance contributions. The antiferromagnetic inner-tetrahedral Heisenberg Hamiltonian density for the spin vectors $\vec S$ and $\vec T$ reads 
\begin{equation} 
V_H=- J_{SS}\sum_{i \neq j=1}^4 \vec S_i\vec S_j
- J_{ST}\sum_{i,j=1}^4 [\vec S_i\vec T_j+\vec T_i\vec S_j] 
- J_{TT}\sum_{i \neq j=1}^4 \vec T_i\vec T_j
\label{mm3}
\end{equation}
where i and j run over the corners of the tetrahedron fig. 1 and the exchange energy densities J can be identified with the couplings introduced in eq. (\ref{n110d})
\begin{eqnarray}  
J_{SS}=J_{TT}=-\frac{1}{2}J_-  \qquad\qquad\qquad J_{ST}=\frac{1}{2}J_+
\label{rot22a}
\end{eqnarray}

Using the commutation relation for the internal spin operators
\begin{eqnarray}  
[\vec S^a_i,\vec S^b_j]=i\epsilon_{abc}\delta_{ij} S^c_i \qquad
[\vec T^a_i,\vec T^b_j]=i\epsilon_{abc}\delta_{ij} T^c_i \qquad
[\vec S^a_i,\vec T^b_j]=0 
\label{mm4xx1}
\end{eqnarray}
one can derive their (internal) time evolution in the Heisenberg picture 
\begin{equation} 
\Lambda^3 \frac{d\vec S_i}{dt}=i[V_H,\vec S_i] \qquad\qquad\qquad 
\Lambda^3 \frac{d\vec T_i}{dt}=i[V_H,\vec T_i]
\label{mm42}
\end{equation}
to obtain
\begin{eqnarray}  
\frac{d\vec S_i}{dt}= \vec S_i^0 \times \sum_{i \neq j=1}^4 [j_{SS}\vec S_j+j_{ST}\vec T_j]+k_{ST}\vec S_i^0\times \vec T_i  \label{eqq22vv} \\
\frac{d\vec T_i}{dt}= \vec T_i^0 \times \sum_{i \neq j=1}^4 [j_{TT}\vec T_j+j_{ST}\vec S_j]+k_{ST}\vec T_i^0\times \vec S_i 
\label{eqq22}
\end{eqnarray}
These equations have been linearized for small displacements $\delta \vec S_i=\vec S_i-\vec S_i^0$ and $\delta \vec T_i=\vec T_i-\vec T_i^0$ of the spin vectors from their ground state positions $\vec S_i^0 =\langle \vec S_i\rangle$ and $\vec T_i^0 =\langle \vec T_i\rangle$ in fig. 3, and the letter $\delta$ has then been left out. Furthermore, we have switched from exchange energy densities 
$J$ to exchange energies 
\begin{eqnarray}  
j=J\Lambda^{-3}
\end{eqnarray} 
Finally, eq. (\ref{eqq22}) includes a contribution, which accounts for the possibility that $\vec S_i \vec T_i$ interact with a different strength than $\vec S_i \vec T_j$ for $j\neq i$, because the internal distance between $\vec S_i$ and $\vec T_i$ is different from the distance between $\vec S_i$ and $\vec T_j$ for $j\neq i$.

The ground state positions are given by
\begin{eqnarray}  
\vec S_1^0=\frac{1}{\sqrt{3}}(-1,-1,-1) \qquad\vec S_2^0=\frac{1}{\sqrt{3}}(-1,+1,+1) \\
\vec S_3^0=\frac{1}{\sqrt{3}}(+1,-1,+1) \qquad\vec S_4^0=\frac{1}{\sqrt{3}}(+1,+1,-1) 
\label{eqq23}
\end{eqnarray}
and $\vec T_i^0=\pm\vec S_i^0$ depending on whether one is analyzing the parallel or anti-parallel configuration fig. 3. An overall normalization factor N of the vacuum spin vectors was put to 1, because it does not influence the eigenfrequencies and mixing angles. More precisely, it can be absorbed by a redefinition of the couplings $j\rightarrow jN$.

So everything fixed now to solve the differential equations (\ref{eqq22})? Not quite, because for a frustrated, i.e. non-collinear ground state configuration like fig. 1 it is necessary to go beyond the collinear spin wave analysis and transform to a rotating frame with the z-axis pointing along the local spin direction\cite{dmz1,dmz2}. Applied to the present case this procedure modifies eqs. (\ref{eqq22}) in such a way that
\begin{eqnarray}  
\frac{d\overrightarrow{U_iS_i}}{dt}= \overrightarrow{U_iS_i} \times \sum_{i \neq j=1}^4 [j_{SS}\overrightarrow{U_jS_j}+j_{ST}\overrightarrow{U_jT_j}] +k_{ST}\overrightarrow{U_iS_i}\times  \overrightarrow{U_iT_i} \label{eqq24vv} \\
\frac{d\overrightarrow{U_iT_i}}{dt}= \overrightarrow{U_iT_i} \times \sum_{i \neq j=1}^4 [j_{TT}\overrightarrow{U_jT_j}+j_{ST}\overrightarrow{U_jS_j}] +k_{ST}\overrightarrow{U_iT_i}\times \overrightarrow{U_iS_i}
\label{eqq24}
\end{eqnarray}
where $U_i$ are diagonal $3\times 3$ matrices which act on the vector components of the spin vectors
\begin{eqnarray}  
U_1=D(1,1,1) \quad U_2=D(1,-1,-1) \quad
U_3=D(-1,1,-1) \quad U_4=D(-1,-1,1) 
\label{eqq25}
\end{eqnarray}

The resulting 24$\times$24 matrix has to be diagonalized in order to obtain the 24 eigenfrequencies $\omega$. Due to the Shubnikov symmetry of the system the corresponding eigenstates can be arranged into 6 singlets and 6 triplets as in eq. (\ref{eq833hg}), i.e. as leptons and quarks. Each triplet consists of 3 states with degenerate eigenvalues, because the Shubnikov symmetry $A_4+S(S_4-A_4)$ is unbroken. 

The result of the diagonalization procedure gives the following non-vanishing masses / eigenfrequencies consisting of 2 singlets
\begin{eqnarray}
\omega (\mu) = -\omega (\tau)&=& \pm(6 j_{ST}+ 2 k_{ST})
\label{mm15f}
\end{eqnarray} 
and 4 triplets 
\begin{eqnarray}
\omega (c) = -\omega (t) &=&    \sqrt{j_1^2 + \Delta j_2 }    
\label{mm15ff44}
\\
\omega (s) = -\omega (b) &=& \sqrt{j_1^2 -  \Delta j_2 }   
\label{mm15ff}
\end{eqnarray} 
where $j_1^2$, $\Delta$ and $j_2$ are abbreviations for
\begin{eqnarray}
j_1^2&=&-8(j_{SS}^2+j_{TT}^2)-10j_{ST}^2-2k_{ST}^2-2(j_{SS}+j_{TT})(6j_{ST}+2k_{ST})
-2j_{ST}k_{ST} \nonumber \\
\Delta^2&=&4(j_{SS}-j_{TT})^2+(j_{ST}-k_{ST})^2 \nonumber \\
j_2&=&4(j_{SS}+j_{TT})+6 j_{ST}+ 2 k_{ST}
\end{eqnarray} 
The corresponding eigenvectors are not given because the formulas are too cumbersome to be presented here.

At this stage, apart from these 14 modes there are 10 zero modes (4 singlets and 2 triplets), which can be attributed to the 3 neutrinos, the electron and the up- and down-quark.
Using $j_{SS}=j_{TT}$ obtained in eq. (\ref{rot22a}) from the heuristic deduction via the generalized NJL model the quark mass eigenvalues simplify to
\begin{eqnarray}
\omega (c) = -\omega (t) &=&4j_{SS} +4j_{ST} 
\label{mm15ff44g}
\\
\omega (s) = -\omega (b) &=&4j_{SS}+2 j_{ST} +2k_{ST}
\label{mm15ffg}
\end{eqnarray} 


One concludes that considering only the anti-ferromagnetic Heisenberg contributions eqs. (\ref{mm3}) and (\ref{eqq22}) leads to a stronger degeneracy than dictated by the Shubnikov symmetry alone. In fact, this does not only concern the zero modes, because according to (\ref{mm15f})-(\ref{mm15ff}) the quarks and leptons of the second and third family have the same mass.

In the next section effects of inter-tetrahedral interactions will be included to partially lift this degeneracy between the families and in particular to shift the masses of the third family to larger values. Most prominently, the top quark mass will be equipped with a contribution of order $\Lambda_F$. Afterwards, torsion and anisotropic corrections will be included. They are tiny effects and cannot be attributed to a SM piece of the interactions like the contributions discussed so far. However, they are needed, because they are responsible for the light quark and lepton masses, and in particular for the neutrino masses and mixings. Hence they will remove the 'accidental' degeneracies which one obtains if one only considers the inner-tetrahedral Heisenberg exchange contributions to the eigenfrequencies.

What can be done already at this point is to take the formulas (\ref{mm15f})-(\ref{mm15ff}) for the second family and to adopt them to the known mass values of the muon, the charmed and the strange quark. It is advisable to take the running values of the masses at order 1 TeV\cite{runningmass} and try to determine from that the values for the internal exchange couplings. A closer look at eqs. (\ref{mm15f})-(\ref{mm15ff}) reveals that there is no simple solution with small or antiferromagnetic (i.e. negative) value of $k_{ST}$. Instead, one obtains
\begin{eqnarray}
j_{ST}\approx -0.12 \ \text{GeV} \qquad j_{TT}=j_{SS}\approx -0.12 \ \text{GeV} \qquad k_{ST} \approx 0.32  \ \text{GeV}
\label{mm15ffff}
\end{eqnarray} 
and concludes that the coupling strengths are $\leq 1$ GeV with a ferromagnetic coupling $k_{ST}$ of adjacent spin vectors $\vec S_i$ and $\vec T_i$, while the interactions with $i\neq j$ are somewhat smaller in magnitude and anti-ferromagnetic. This is precisely, what was anticipated for the vacuum configuration eq. (\ref{nba77p}). 

\section{Top Quark Mass from Dzyaloshinskii Moriya Interactions}

As discussed before, the {\it inter}-tetrahedral interactions should yield the leading SSB contributions to the fermion masses. To account for these contributions one has to include a sum over neighboring tetrahedrons in the spin Hamiltonian and add the effect of the corresponding interactions to the e.o.m. Since I do not have the resources to treat these many terms properly the following trick will be used to approximately solve the problem: the effects of all other tetrahedrons on a given tetrahedron fig. 3 will be subsumed as an effective contribution generated by the internal spins. The idea is that this effective contribution can be attributed to the gauge transformation eq. (\ref{hipoga}) which transfers the $\vec\pi$-field from the Higgs sector to the W-boson mass term. As well known such a transformation modifies the W-field by
\begin{eqnarray}
\vec W_\mu \rightarrow  \vec W_\mu - \frac{1}{g\Lambda_F}\partial_\mu \vec\pi - \frac{1}{\Lambda_F}\vec\pi \times \vec W_\mu
\label{mm19u1}
\end{eqnarray} 
Thus, while the bilinear terms in $\vec\pi$ disappear from the Higgs potential eq. (\ref{hipo}), they re-appear in the W-mass term of the Lagrangian. Furthermore, their sign is such that the antiferromagnetic $\vec\pi - \vec\pi$ coupling at high energies gets transformed into a ferromagnetic interaction plus an additional term which is due to the non-abelian nature of the gauge transformation. All in all
\begin{equation} 
\frac{1}{2} m_W^2\vec W_\mu \vec W^\mu \rightarrow \frac{\Lambda_F}{8}  \sum_{i,j=1}^4 [ \vec S_i\vec S_j  +i (\vec S_i\times \vec D_{ij}) \vec S_j +c.j. ] 
\label{mm3dm}
\end{equation}
has to be added to the Heisenberg Hamiltonian (\ref{mm3}). To obtain this result the tree level relation $g=2m_W/\Lambda_F$ has been used. Note that because of the $SU(2)_L$ interactions only terms involving $\vec S$ and its charge conjugate (c.j.) $\vec V$ appear, as defined in (\ref{njl4b87}), in accordance with the V-A structure of the weak interactions.
This is in agreement with ref. \cite{laa} where it has been shown that after the formation of the chiral tetrahedrons the internal spin $SU(2)$ becomes a symmetry involving only left handed particles, $SU(2)_L$.

The additional term in eq. (\ref{mm3dm}) involving the cross product can be interpreted as a Dzyaloshinskii-Moriya component\cite{dmgut}, a contribution which in solid state physics is sometimes used to describe leading anisotropic corrections to the ordinary Heisenberg equations of motion. Quite in general, such a component stands for a tendency to form a rotational structure (instead of the ordinary ferromagnetic alignment of neighboring tetrahedrons depicted in fig. 2) simply because the DM-term tends to rotate the spin vectors instead of aligning them. In the present framework it was deduced as a consequence of the gauge transformation eq. (\ref{hipoga}). Therefore the DM-term can be interpreted quite naturally, namely by that the $SU(2)_L$ gauge fields induce a curvature of the fiber bundle formed by the system of all tetrahedrons, and the DM-term simply takes care of this curvature effect to effectively maintain the aligned structure.

In general, a DM-component to the Heisenberg Hamiltonian has a complicated coupling structure involving vector couplings $\vec D_{ij}$, as given in eq. (\ref{mm3dm}). However, in the case at hand the couplings $\vec D_{ij}$ are fixed by 2 symmetry requirements, namely that the term must be $SU(2)_L$ gauge invariant and that it must respect the $A_4+S(S_4-A_4)$ Shubnikov symmetry. While the former fixes the modulus of the DM coupling strength relative to the Heisenberg term in eq. (\ref{mm3dm}), the latter forces the direction of $\vec D_{ij}$ to be $\vec S^0_k-\vec S^0_l$\cite{dmgut}, where $kl\neq ij$ is chosen such that the sign of the permutation (ijkl) of (1234) is positive\cite{dmgut}.

The equations of motion of the spin vectors (\ref{eqq22vv}) are supplemented by the inter-tetrahedral contribution (\ref{mm3dm}) in the following way:
\begin{eqnarray}  
\frac{d\vec S_i}{dt}= \frac{1}{4} \Lambda_F \{ \vec S_i \times ( [ 1+ i  \sum_{j}\vec D_{ij} \times ]   \vec S_j ) \}
\label{eqq212}
\end{eqnarray}
These equations modify only the 12 equations (\ref{eqq22vv}) for $\vec S_i$ leaving the equations for the $SU(2)_R$ vibrators $\vec T_i$ (\ref{eqq22}) untouched. Their net effect is to give a quasi-democratic type of contribution $\sim \Lambda_F$ to the mass matrix, in the sense that each entry of the 12$\times$12 eigenmatrix for $\vec S_i$ gets a contribution involving $\Lambda_F$. Evaluating the eigenvalues, this equips the top quark triplet with a mass 
\begin{eqnarray}  
m_t=N \Lambda_F=\frac{\Lambda_F}{2g}=\frac{\Lambda_F^2}{4m_W} \approx 190 GeV
\label{eqqmt}
\end{eqnarray}
while leaving the other quark and lepton masses unchanged.

An overall factor $N$ has been included which arises from the normalization of the spin vectors and was left out in eqs. (\ref{eqq23}) and (\ref{mm15f})-(\ref{mm15ffg}) because it was absorbed in the redefinition of the Heisenberg couplings. This is not possible in the case at hand, because the coupling is given in terms of the predefined quantity $\Lambda_F$. $N$ can be shown to be related to the $SU(2)_L$ gauge coupling g via $2gN=1$. 

Now that we have established a top quark mass of order $\Lambda_F$ one may ask whether there are contributions to the masses of the $\tau$-lepton and the b-quark from the inter-tetrahedral couplings as well. Such contributions are strongly desired, because one would like to get rid of the relations $m_b=m_s$ and $m_\tau=m_\mu$ following from eqs. (\ref{mm15f}) and (\ref{mm15ff}). Unfortunately, if one sticks to (\ref{mm3dm}) and (\ref{eqq212}), the answer to that question is no. These equations only equip the top quark with a mass.

However, one should keep in mind that (\ref{mm3dm}) and (\ref{eqq212}) represent a rather crude approximation to the inter-tetrahedral interaction effects. One may analyze several possible modifications on whether they lead to the desired effect. For example, one may think of a contribution due to $\gamma$-Z mixing. As well known, this mixing leads to a modification of the W-mass term in the form $\sim m_Z^2 Z_\mu Z^\mu$ where $m_Z:=m_W/\cos \theta_W$ and $Z:=W_z\cos\theta_W-B\sin\theta_W$ and $B$ is the SM $U(1)_Y$ gauge field. One obviously has $m_Z^2 Z_\mu Z^\mu=m_W^2 W_{z\mu} W_z^\mu+O(B)$. In effect the spin-spin interaction eq. (\ref{mm3dm}) is not modified, the physical reason being that the $U(1)_Y$ terms are isospin blind and therefore do not modify directions in the internal space. 

Another possibility are mixings among the various fields in eq. (\ref{njl09w}). In fact, the quantum numbers of the Higgs allow for a mixing with the z-component of the iso-vector $\vec v$. Similarly, the pseudo-scalar combination $\eta\sim \bar\psi i\gamma_5 \psi$ can mix with $\pi_z$. However, according to my analysis there is again no effect on the quark and lepton masses.

Finally there may be an admixture of a right handed current in such a way that $\vec S$ in eq. (\ref{mm3dm}) should effectively be replaced by 
\begin{eqnarray}  
\vec S \rightarrow \vec S+\alpha \vec T
\label{eqqrh}
\end{eqnarray}
It is true that there are rather strong restrictions to the presence of right handed currents and on the value of $\alpha$, but effects in the percent range are still allowed \cite{nesti,moha2}.
I am not claiming that they are actually present, in particular, because a right-handed W-boson contradicts the arguments advocated in ref. \cite{laa}, that the handedness of the tetrahedral structure fig. 1 should make any weak interaction left-handed. I am just considering them for the sake of having a definite example, because such terms turn out to have the nice feature that the mass contribution to the top quark essentially remains unchanged while new contributions to the b- and $\tau$-mass are generated of order $\sim\alpha\Lambda_F$. More precisely, the contribution from this source is
\begin{eqnarray}  
m_b=m_\tau= \frac{3\alpha\Lambda_F^2}{16m_W} 
\label{eqqrh6}
\end{eqnarray}
which interestingly is in accord with the Georgi-Jarlskog mass relations\cite{georgi}. Note that all lighter quark and lepton masses are not modified by the $\alpha\vec T$ term in eq. (\ref{eqqrh}). 

\section{Neutrino Masses and the Conservation of internal Angular Momentum}

The calculations presented so far are based on a certain interpretation of the Standard Model Higgs mechanism - a rather intuitive interpretation, if one is willing to accept that the form of the quark and lepton spectrum is due to a discrete internal tetrahedral structure. As was shown above it is possible to identify the terms in the SM Lagrangian responsible for the internal spin interactions, namely the quadratic part of the Higgs potential and the W-mass term. However, at this point 10 of the 24 possible excitations (the neutrinos, the electron and up and down quarks) are still without a mass. 

As long as these 4 singlets and 2 triplets remain massless, i.e. constant, non-vibrating modes, it is no use trying to calculate the CKM matrix elements or even the mixing angles in the neutrino sector. To get rid of those degeneracies one has to relax a condition inherent in the classical Heisenberg model namely that the internal magnetic moments may be treated as classical 3-dimensional vector spins of fixed length. This condition is destroyed by quantum fluctuations in the quantum Heisenberg model and on the classical level by allowing for torsional vibrations. 
Although these (tiny) torsional effects have no counterpart in the SM Lagrangian it can be shown that the leading up-quark, down-quark and electron mass contributions are provided by isotropic torsional interactions of the internal spin vectors while the differences in the neutrino masses can be attributed to anisotropic effects within the torsional couplings.

To start with I am now going to write down the most general form of these torsional interactions.
As argued above, torsion is not strictly forbidden in the system under consideration, and for the case of only one spin vector is simply induced by a contribution of the form $d\vec S /dt \sim \vec S$ to its time variation. In the case at hand with 8 spin vectors $\vec S_i$ and $\vec T_i$ its main effect is to allow vibrations along the local z-directions and thus lift the degeneracies of all zero modes. The terms supplementing the Heisenberg e.o.m. (\ref{eqq22vv}) and (\ref{eqq22}) are
\begin{eqnarray}  
\frac{d\vec S_i}{dt}= i e_{SS} \vec S_i +if_{SS} \sum_{j\neq i} \vec S_j+ie_{ST} \vec T_i +if_{ST} \sum_{j\neq i} \vec T_j  \label{eqq22ii} \\
\frac{d\vec T_i}{dt}= i e_{TT} \vec T_i +if_{TT} \sum_{j\neq i} \vec T_j+ie_{ST} \vec S_i +if_{ST} \sum_{j\neq i} \vec S_j 
\label{eqq22iii}
\end{eqnarray}
where e and f are the torsion coupling strengths, whose values are assumed to be small compared to the exchange couplings considered so far. More precisely one has the natural hierarchy
\begin{eqnarray}
e,f \sim O(MeV) \ll j,k \sim O(GeV) \ll \Lambda_F \alpha \ll \Lambda_F  
\label{eqq2113}
\end{eqnarray}
so that the torsional couplings can indeed be expected to provide for the electron and the up- and down-quark mass with $m_{e,u,d}/m_{\mu,s,c} \sim 10^{-2}$. As will be shown in the next section neutrino masses are due to still smaller anisotropy effects among the torsional couplings.

The contributions (\ref{eqq22ii}) and (\ref{eqq22iii}) can be incorporated in the full 24$\times$24 eigenvalue problem to yield
\begin{eqnarray}
m_e &=& e_{SS}-e_{ST}+3f_{SS}-3f_{ST} \label{mas86} \\
m_\mu &=& 6 j_{ST}+ 2 k_{ST}+O(m_e) \\
m_\tau  &=& \frac{3\alpha\Lambda_F^2}{16m_W} + O(m_\mu) 
\end{eqnarray}
\begin{eqnarray}
m_u &=& e_{SS}+e_{ST}-f_{SS}-f_{ST} \\
m_c &=& 4j_{SS} +4j_{ST} +O(m_u)  \\
m_t &=& \frac{\Lambda_F^2}{4m_W} + O(m_c)  
\end{eqnarray}
\begin{eqnarray}
m_d &=& e_{SS}-e_{ST}-f_{SS}+f_{ST}  \\
m_s &=& 4j_{SS}+2 j_{ST} +2k_{ST}+O(m_d)  \\
m_b &=& \frac{3\alpha\Lambda_F^2}{16m_W} + O(m_s)  
\end{eqnarray}
\begin{eqnarray}
m(\nu_e) &=& m(\nu_\mu) = m(\nu_\tau) =e_{SS}+e_{ST}+3f_{SS}+3f_{ST} 
\label{mas8}
\end{eqnarray}
For simpicity and to be in accord with the NJL expectation $j_{SS}=j_{TT}$ eq. (\ref{rot22a}) I have set $e_{SS}=e_{TT}$ and $f_{SS}=f_{TT}$.

It is interesting to note that eq. (\ref{mas8}) seems to indicate that the neutrinos acquire a mass, too - in fact the same mass for all neutrino species. Actually, the quantity on the right hand side of (\ref{mas8}) deserves special attention, because it governs the evolution of the total internal angular momentum $\vec \Sigma:=\sum_i (\vec S_i+\vec T_i)$ as can be seen by adding up all contributions eqs. (\ref{eqq22ii}) and (\ref{eqq22iii})
\begin{eqnarray}  
\frac{d \vec \Sigma}{dt}\equiv \frac{d\sum_{i=1}^4 (\vec S_i+\vec T_i)}{dt}= i (e_{SS}+e_{ST}+3f_{SS}+3f_{ST}) \sum_{i=1}^4 (\vec S_i+\vec T_i)
\label{eqsu8}
\end{eqnarray}
It should be noted that the Heisenberg interactions (\ref{mm3}) as well as the DM term (\ref{eqq212}) fulfill $d\vec\Sigma /dt=0$ and thus do not contribute to eq. (\ref{eqsu8}).

Comparing (\ref{mas8}) and (\ref{eqsu8}) one concludes, that the 3 neutrino eigenstates correspond to the vibrations of the 3 components of $\vec \Sigma$. Whenever the total internal angular momentum is conserved, i.e. $\vec \Sigma$ is independent of the (internal) time, the neutrino masses will strictly  vanish, while non-zero neutrino masses are a signal for non-conservation of $\vec \Sigma$. 
In the framework of Noether's Theorem $\vec \Sigma$ is the conserved charge related to the internal rotational symmetry, and the neutrino states $\nu_e$, $\nu_\mu$ and $\nu_\tau$ are related to the breaking of this continuous symmetry. Applying Goldstone's theorem to the internal dynamics yields 3 internal massless magnon excitations, in a similar way, as magnons are obtained as Goldstone bosons of the broken rotational symmetry in ordinary magnetic systems.

Goldstone bosons? This sounds strange in view of the fact that neutrinos are fermions. The point is that one has to distinguish the dynamics in internal from that in physical space. In physical space the neutrinos are fermions, but in internal space they are described by (bosonic) excitations of the internal angular momentum $\vec S +\vec T$ which is the conserved quantity associated with the internal rotational symmetry.

In principle, the general solution to the eigenproblem does not only give the eigenvalues (\ref{mas86})-(\ref{mas8}) but via the corresponding eigenvectors can also be used to accommodate all physical quark and lepton mixing parameters. Due to lack of resources I have to leave the determination of the quark mixings 
to future work. To get some understanding of the physics of the mixing I will now concentrate on the lepton sector of the theory, which is somewhat easier to handle than the full 24$\times$24 problem. To obtain non-vanishing differences between the masses of the neutrino species additional interaction terms violating internal angular momentum conservation have to be introduced. Afterwards, the method how to determine the PMNS mixing matrices from these interactions within the present framework will be desribed in detail.


\section{(Not just) a Toy Model for Leptons}


The mass problem for the leptons alone can be approximately reduced from the tetrahedral configuration to the simple 1-dimensional structure depicted in fig. 4. As Eq. (\ref{mm15f}) indicates, the lepton masses do not depend on the couplings $J_{SS}$ and $J_{TT}$ but only on $J_{ST}$. 
As a matter of fact using some simple matrix algebra manipulations it may be shown that the Heisenberg contributions to the Shubnikov singlets (leptons) can effectively be obtained from the configuration of fig. 4 with only 2 internal spin vectors $\vec S$ and $\vec T$ (instead of 8) which in the ground state point in the z-directions 
\begin{equation} 
\langle\vec S\rangle=\langle\vec T\rangle=(0,0,1)
\label{mm5}
\end{equation}

To analyze the behavior of this system, I will start with the Heisenberg part of the interactions
\begin{equation} 
V_H=-J \vec S \vec T
\label{mm655}
\end{equation}
where $J$ is identical to $4J_{ST}$ used in the last section. The factor of $4$ is a geometrical factor arising from the reduction of the tetrahedral configuration.

\begin{figure}
\begin{center}
\epsfig{file=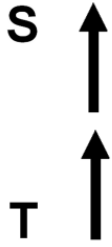,height=3.5cm}
\caption{The local ground state in the model for leptons, where the spin vectors $\vec S$ and $\vec T$ point in the z-direction. The parallel configuration is depicted here instead of the anti-parallel one in fig. 3. If only a Heisenberg interaction of the form (\ref{mm655}) is included the combination $\vec S +\vec T$ remains static, thus giving rise to 3 zero modes corresponding to the 3 neutrinos. A fourth zero-mode appears in this limit, because the spin vectors will rotate only in the transversal $(x,y)$-plane, leaving all torsional vibrations (i.e. in the z-direction) as zero modes. If further interactions are added to the Heisenberg term, all 6 eigenmodes receive non-vanishing values.}
\nonumber
\end{center}
\end{figure}

The 6 d.o.f. of this system of 2 internal chiral spin vectors lead to 6 eigenmodes, and we are now going to show how the lepton masses and mixings, in particular the tiny neutrino masses may arise. The symmetry group of the ground state fig. 4 is $\{1,SR\}$, whose only non-trivial element is a reflection R at the x-y-plane followed by an internal time reversal S. 
This group has only singlet representations\cite{shub}, according to which the 6 eigenmodes will be classified.

\begin{table}
\begin{center}
\begin{tabular}{|l|c|c|c|c|c|c|}
\hline
   &$S_x$&$S_y$&$S_z$&$T_x$&$T_y$&$T_z$\\ 
\hline
$S_x$ & $i(\omega_f+v)$ & $j+v$      & $0$       & $i(f-v)$          & $-j-v$       &  $0$  \\
$S_y$ & $-j-v$       & $i(\omega_f+v)$ & $0$      & $j+v$      & $i(f-v)$          &  $0$  \\
$S_z$ & $0$      &  $0$      & $i\omega_f$ & $0$       & $0$      &  $if$  \\
$T_x$ & $i(f-v)$          &  $-j-v$      & $0$      & $i(\omega_f+v)$ & $j+v$      &  $0$  \\ 
$T_y$ & $j+v$      &  $i(f+v)$         & $0$       & $-j-v$       & $i(\omega_f+v)$ & $0$  \\ 
$T_z$ & $0$       & $0$      & $if$          & $0$      & $0$       & $i\omega_f$    \\
\hline
\end{tabular}
\bigskip
\caption{The interaction matrix between the 2 chiral spin vectors $\vec S$ and $\vec T$ of figure 4 giving rise to electron-, muon- and tau-mass. In addition to the Heisenberg interaction eq. (\ref{mm655}) inter-tetrahedral effects ($v$) and a universal torsional coupling ($f$) have been introduced. The neutrinos are still massless at this stage, corresponding to the 3 d.o.f. of $\vec S+\vec T$ which do not vibrate. The abbreviation $\omega_f=\omega -f$ is used.}
\label{tab331}
\end{center}
\end{table}

Adding a small universal torsional coupling $f \ll j$ the time evolution of the spin vectors is given by
\begin{equation} 
\frac{d\vec S}{dt}=j \vec S \times \vec T +f (\vec S -\vec T)\qquad\qquad
\frac{d\vec T}{dt}=j \vec T \times \vec S+f (\vec T-\vec S)
\label{mm7nor}
\end{equation}
One then has to diagonalize the sum of the matrices given in table \ref{tab331} in order to obtain the eigenstates 
\begin{eqnarray}
\nu_e&=&(0, 0, 1, 0, 0, 1) \qquad
e=(0, 0, -1, 0, 0, 1) \nonumber \\
\nu_\mu&=&(0, 1, 0, 0, 1, 0) \qquad 
\mu=(-i, -1, 0, i, 1, 0) \nonumber \\
\nu_\tau&=&(1, 0, 0, 1, 0, 0) \qquad
\tau=(i, -1, 0, -i, 1, 0)
\label{mm23rr}
\end{eqnarray} 
and their masses/eigenfrequencies
\begin{eqnarray}  
\omega(\nu_e)&=&0 \qquad\qquad\qquad
\omega(e)=2f  \nonumber \\
\omega(\nu_\mu)&=&0 \qquad \qquad\qquad
\omega(\mu)=2(j+f) \nonumber \\
\omega(\nu_\tau)&=&0 \qquad \qquad\qquad
\omega(\tau)= 2(2v-j-f) 
\label{eqq2260}
\end{eqnarray}
In eq. (\ref{mm23rr}) a 6-dimensional vector space of eigenvectors has been introduced in which the sum and difference $\vec S \pm \vec T$ are simply given by
\begin{eqnarray}
S_x\pm T_x=\frac{1}{\sqrt{2}}(1,0,0,\pm 1,0,0) \nonumber \\
S_y\pm T_y=\frac{1}{\sqrt{2}}(0,1,0,0,\pm 1,0) \nonumber \\
S_z\pm T_z=\frac{1}{\sqrt{2}}(0,0,1,0,0,\pm 1)
\label{mm23h}
\end{eqnarray} 

\begin{table}
\begin{center}
\begin{tabular}{|l|c|c|c|c|c|c|}
\hline
   &$S_x$&$S_y$&$S_z$&$T_x$&$T_y$&$T_z$\\ 
\hline
$S_x$ & $i e_{xy}$ & $-j_z$& $j_{xy}$ & $if_{xy}$ & $-l_z$&  $-l_{xy}$  \\
$S_y$ & $j_z$ & $ie_{xy}$ & $-j_{xy}$ & $l_z$ & $if_{xy}$&$l_{xy}$  \\
$S_z$ & $-j_{xy}$ &$j_{xy}$ & $ie_z$ & $l_{xy}$ & $-l_{xy}$&  $if_z$  \\
$T_x$ & $if_{xy}$ &$-l_z$& $-l_{xy}$& $ig_{xy}$ & $-m_z$&  $m_{xy}$  \\ 
$T_y$ & $l_z$&$if_{xy}$& $l_{xy}$ & $m_z$& $ig_{xy}$ & $-m_{xy}$  \\ 
$T_z$ & $l_{xy}$& $-l_{xy}$& $if_z$ & $-m_{xy}$& $m_{xy}$& $ig_z$    \\
\hline
\end{tabular}
\bigskip
\caption{The most general correction terms to the matrix in table \ref{tab331}.
}
\label{tab3322}
\end{center}
\end{table}

Furthermore, a contribution $v\sim \alpha \Lambda_F$ reflecting the leading inter-tetrahedral contribution to the $\tau$ mass has been included in table \ref{tab331} and eq. (\ref{eqq2260}) which lifts the degeneracy between muon and $\tau$-lepton.  The natural hierarchy between these couplings can then be invoked to accommodate the lepton masses. The electron mass is naturally small as compared to $m_\mu$ and $m_\tau$ because the electron corresponds to a torsional vibration (of $S_z - T_z$) in the z-direction and its mass gets only torsional contributions $\sim f \approx 0.25$MeV. There is no contribution $\sim j$ to the electron mass because the Heisenberg interaction conserves each spin's fixed length and does not allow spin vibrations in the z-direction. Note the other mode in the z-direction, the one $\sim S_z + T_z$, is to describe the electron neutrino.

The Hamiltonian on which eq. (\ref{mm7nor}) is based conserves the total spin of the system so that the 3 d.o.f. corresponding to this quantity do not vibrate. Comparing (\ref{mm23rr}) with (\ref{mm23h}) one explicitly sees that the neutrino zero modes correspond to the 3 components of the conserved total internal angular momentum $\vec S +\vec T$. As discussed in detail in the last section they can be interpreted as Goldstone modes of the broken internal rotational invariance. In order to obtain non-zero neutrino masses one may add a small contribution $\sim \vec S +\vec T$ to the right hand side of eq. (\ref{mm7nor}). Similar to (\ref{mas8}) this would equip all neutrino species with the same mass.

In order to be more general and obtain different masses for the 3 neutrinos I have written down the most general interaction matrix which violates internal angular momentum conservation  and includes anisotropic and torsional forces in table \ref{tab3322}. Compared to the leading terms in table \ref{tab331} the new contributions must be tiny, as are the neutrino masses. More precisely, they should be of order at most $ O(m_\nu /m_e)\approx 10^{-7}$.

It seems clear that corrections of such minuteness are difficult to handle quantitatively. Nevertheless, the present approach allows to analyze the question from which of the various sources appearing in table \ref{tab3322} the observed neutrino masses and mixings\cite{nuemix} actually arise. For example, an inverted hierarchy of neutrino masses which seems to be slightly favored by the present data can be accommodated quite easily, and measured values of the mixing angles will further determine the coupling parameters in table \ref{tab3322}.  

A complete numerical analysis of the whole parameter space is not undertaken in the present study. Rather, one realistic example will be discussed to see whether the neutrino measured parameters can be reproduced. Namely, a simple solution to the case of the inverted hierarchy is obtained by putting $j_{xy}=l_{xy}=m_{xy}\equiv \delta_j$ and $f_{xy}\equiv \delta_e$ and all other parameters in table \ref{tab3322} to zero. The result for the masses then is
\begin{eqnarray}
\omega(\nu_e)&=&-\delta_e \qquad \qquad\qquad\qquad \qquad\qquad
\omega(e)=2f   \\
\omega(\nu_\mu)&=&-\frac{1}{2}\{\delta_e+\sqrt{\delta_e^2+32\delta_j^2}\} \qquad \qquad
\omega(\mu)=2(j+f)+\delta_e \nonumber \\
\omega(\nu_\tau)&=& -\frac{1}{2}\{\delta_e-\sqrt{\delta_e^2+32\delta_j^2}\} \qquad\qquad
\omega(\tau)= 2(2v+j-f)+\delta_e \nonumber
\label{mm15}
\end{eqnarray} 
Assuming $\delta_e \gg \delta_j$ one can indeed reproduce the 'inverted hierarchy' of neutrino masses. To be definite we choose the values $\delta_j=0.002$eV, $\delta_e=0.05$eV, $f=0.25$MeV $j=50$MeV $v=0.4$GeV in order to reproduce the current data, i.e.
\begin{eqnarray}
m_2^2-m_1^2=0.000063\ eV^2 \\
m_3^2-\frac{1}{2}(m_1^2+m_2^2)=-0.0025\ eV^2
\label{mm1580}
\end{eqnarray} 
Next one has to check whether the mixing angles for the neutrino come out right. This is then the appropriate place to discuss the general strategy how mixing matrices can be obtained in the present framework. To determine the CKM quark mixing elements the eigenvectors of a complicated 24$\times$24 problem have to be fixed. For the case of leptons with the 6$\times$6 matrix in tables \ref{tab331} and \ref{tab3322} the calculation of eigenvectors and PMNS mixing matrix elements is a relatively simple exercise. 
 
One just has to remember that the mixing matrix is defined as the unitary transition matrix $U_N=(U_{N\alpha a})$ between the mass eigenstates $\nu_\alpha, \alpha=e,\mu,\tau$, and the weak interaction eigenstates $\nu_a, a=1,2,3$:
\begin{eqnarray}
\nu_\alpha = \sum_{a} U_{N\alpha a}\nu_a \longleftrightarrow \nu_a = \sum_{\alpha} U^*_{N\alpha a}\nu_\alpha
\label{am164}
\end{eqnarray} 
The corresponding matrix for the charged leptons will be denoted by $U_E$.
The PMNS matrix is then given by the product
\begin{eqnarray}
V_{PMNS}=U_N^{\dagger}U_E
=\begin{bmatrix}
\langle \nu_e | \nu_1 \rangle & \langle \nu_e | \nu_2 \rangle & \langle \nu_e | \nu_3 \rangle\\
\langle \nu_\mu | \nu_1 \rangle & \langle \nu_\mu | \nu_2 \rangle & \langle \nu_\mu | \nu_3 \rangle\\
\langle \nu_\tau | \nu_1 \rangle & \langle \nu_\tau | \nu_2 \rangle & \langle \nu_\tau | \nu_3 \rangle
\end{bmatrix}
\begin{bmatrix}
\langle e_1 | e \rangle & \langle e_1 | \mu \rangle & \langle e_1 | \tau \rangle\\
\langle e_2 | e \rangle & \langle e_2 | \mu \rangle & \langle e_2 | \tau \rangle\\
\langle e_3 | e \rangle & \langle e_3 | \mu \rangle & \langle e_3 | \tau \rangle
\end{bmatrix}
\label{mm23cc}
\end{eqnarray}
where a mnemonic notation $\langle | \rangle$ has been used for the matrix elements of $U_N$ and $U_E$.

In the case at hand the mass eigenstates can be directly identified with the eigenvectors of the given eigenvalue problem tables \ref{tab331} and \ref{tab3322}. For the special values given after eq. (\ref{mm15}) one obtains
\begin{eqnarray}
\nu_e&=&(-0.034 - 0.49i, 0.49, 0.072 - 0.076i, -0.034 - 0.49i, 0.49, 0.072 - 0.076i) \nonumber \\ 
e&=&(0,0,-0.707,0,0,0.707) \nonumber \\
\nu_\mu&=&(0.078 + 0.47i, 0.476, -0.166 - 0.14i, 0.078 + 0.47i, 0.476, -0.166 - 0.14i) \nonumber \\ 
\mu &=&(0.500i, 0.500, 0, -0.500i, -0.500, 0) \nonumber \\
\nu_\tau &=& (0.066 + 0.158i, 0.066 - 0.158i, 0.665, 0.066 + 0.158i, 0.066 - 0.158i, 0.665)\nonumber \\
\tau&=&(0.500i, -0.500, 0, -0.500i, 0.500, 0)
\label{mm23}
\end{eqnarray} 
where the 6-dimensional notation of eqs. (\ref{mm23h}) has been used. 

According to eq. (\ref{njl4baa}) the isospin vector $\vec S$ gives the left-handed particle contribution, while $\vec T$ does the same for the righthanded anti-particles. Therefore the transformation from the neutrino and charged lepton mass eigenstates to their interaction eigenstates is given by the projection of the eigenvectors (\ref{mm23}) to $\vec S$ and $\vec T$, respectively. 
Carrying out these operations one obtains
\begin{eqnarray}
V_{PMNS}=
\begin{pmatrix}
-0.627 + 0.144 i & 0.158 + 0.542 i & 0.141 + 0.500 i \\
0.012 + 0.311 i & 0.410 + 0.158 i & -0.831 - 0.141 i \\
0.047 + 0.697 i & -0.698+0.001i & -0.101 + 0.108 i 
\end{pmatrix}
\label{ccjj1}
\end{eqnarray}
There is a CP violating effect in this matrix, because the Jarlskog invariant\cite{jarl} $J_{CP}$ is non-zero and given by $J_{CP}=0.0222$.

The point to note here is that large mixing angles for the PMNS matrix appear quite naturally, because the neutrino states lie near the total angular momentum $\vec \Sigma=\vec S+\vec T$ which is quite far away from the projection operator $\vec S$. In contrast, due to the dominance of the inter-tetrahedral interactions eq. (\ref{mm3dm}) the top quark state lies very close to the projection vector $\vec S$. In addition, this dominance will align all other quark flavors, thus forcing their CKM elements to small values.

On the quantitative side the result eq. (\ref{ccjj1}) unfortunately does not really reproduce the observed mixing angles for the neutrinos\cite{nuemix,altamix}.
A complete scan of the full parameter space seems unavoidable and should eventually better be carried out for the complete 24$\times$24 eigenvalue problem. This effort will be undertaken in future work. 

\section{Conclusions}


In the preceeding sections a microscopic model for the SM Higgs mechanism has been applied to determine the quark and lepton masses and mixing angles. A discrete tetrahedral structure within a dynamical internal space has managed to fill the gap between the phenomenological hierarchy of mass scales. The underlying physical picture is that the universe resembles a huge crystal of internal molecules, each 'molecule' of tetrahedral form and arranged in such a way that certain symmetries are (spontaneously) broken. For such a model to be consistent, a (6+1)-dimensional space time has been introduced in ref.\cite{laa}, i.e. the 'molecules' extend to 3 internal dimensions orthogonal to physical space, and they interact via a (6+1)-dimensional QED featuring the necessary 'iso-magnetic' forces. The present-day structure of the universe is that of a (3+1)-dimensional 'surface crystal' built from an infinite set of tetrahedrons and living within the (6+1)-dimensional space time. For reasons of symmetry, no growth of the crystal is allowed into the internal dimensions.

The strong correlations within this system provide for the Higgs particle and the weak vector bosons as bound states. 
Furthermore, internal spin excitations turn out to generate the correct quark and lepton spectrum. Then, it happens that an excitation in one internal tetrahedron is able to excite an excitation in the neighboring internal space and thus can travel as a quasi-particle through Minkowski space with a certain wave vector $\vec k$ which is to be interpreted as the physical momentum of the quark or lepton. 

In this model, the SM symmetry breaking can be understood to proceed in 2 steps: 
\begin{itemize}
\item the formation of a tetrahedron due to a new internal interaction force, which is 'antiferromagnetic' at small distances and leads to a frustrated configuration of isospin vectors fig. 1. 
The frustrated tetrahedron breaks the internal spin-SU(2) as well as internal parity and time reversal to the Shubnikov group $A_4+S (S_4 - A_4)$. This symmetry breaking is not spontaneous but arises from the internal antiferromagnetic exchange interaction which avoids parallel spin states. The resulting local ground state fig. 1 is a chiral configuration, i.e. it violates internal and (as shown in ref.\cite{laa}) external parity, and the whole system is left $SU(2)_L$-symmetric in the following sense:
\item each local tetrahedral ground state can rotate independently of the others, i.e. it can freely rotate as a rigid body over its base point in Minkowski space, and this rotational symmetry of the rigid chiral spin vector system corresponds to a $SO(3)$ symmetry group, whose covering group is taken to define $SU(2)_L$. As a matter of fact it is a local symmetry, because the rotation can be different for tetrahedrons over different base points. Furthermore, due to the $V-A$ structure of the interactions induced by the chiral tetrahedral structure, it is a symmetry involving only left handed particles\cite{laa}. At large distances of the order of the Fermi scale the new interactions are (internally) ferromagnetic in nature and give rise to the global ferromagnetic order shown in fig. 2. Finally, the non-vanishing vev for the Higgs field $\sim \bar \psi \psi$ is due to a pairing mechanism as described in ref.\cite{laa}. 
\end{itemize}

In order to analyze the mass problem of the fermions within this model, the most general $SU(2)_L\times SU(2)_R$ invariant NJL Lagrangian has been used\cite{njl2} as an effective approximation. Afterwards, a Heisenberg Hamiltonian for the internal spin vector interactions has been derived from that Lagrangian. This is justified because at the stage when the internal tetrahedron is formed chiral symmetry is still valid, so that one can describe the internal spin vibrations in terms of the chiral spin vectors $\vec S$ and $\vec T$.
Concerning the 'ferromagnetic' attraction  between different tetrahedrons at large distances responsible for the SSB, it was noticed in section 5 that the gauge structure enforces an additional term which resembles the so called Dzyaloshinskii-Moriya interactions\cite{dmgut} of solid state physics. This term was interpreted quite naturally as a curvature effect of the $SU(2)_L$ gauge fields induced in the fiber bundle formed by the system of all tetrahedrons. The DM term simply takes care of this curvature to effectively maintain the 'ferromagnetic' order fig. 2.

Based on the prescribed model expressions for the quark and lepton mass spectrum were derived. It turned out that the extreme hierarchy in this spectrum can be attributed to the fact that the masses of different fermions get contributions from physically different sources, namely
\begin{itemize}
\item the top mass is dominated by a contribution of order $\Lambda_F$ which stems from the SSB {\it inter}-tetrahedral DM interactions. Physically it arises because the top quark corresponds to the 3 eigenmodes which 'disturb' the global ground state in the strongest possible way. This disturbance is also responsible for the hierarchy observed in the CKM matrix elements.
\item strange-, charm- and muon-mass are dominated by antiferromagnetic exchange couplings within one tetrahedron, and thus can be obtained from the {\it inner}-tetrahedral Heisenberg exchange couplings alone. 
\item down-quark, up-quark and electron get their relatively small masses from energetically favored torsion contributions, which only concern 'radial' excitations of the internal spin vectors.
\item neutrino masses are protected by symmetry, because they correspond to vibrations in the valleys of the potential where all Heisenberg and even most of the torsional energy contributions vanish. This was shown to be related to a Goldstone effect associated to the breaking of the internal rotational SO(3) to the tetrahedral symmetry. While in ordinary ferromagnets after magnetization a U(1) symmetry about the z-axis survives, in the given frustrated configuration fig. 1 all three SO(3) generators give rise to Goldstone bosons, to be identified as the internal magnons corresponding to the 3 neutrino species. 
\end{itemize}

Furthermore, the question of quark and lepton mixing was considered, albeit not in a very elaborate way.
The quark mixing is a complicated 24$\times$24 eigenvector problem with many parameters and a detailed analysis therefore postponed to future work. An attempt to determine the lepton mixing parameters was made. It turned out that the phenomenological values for the PMNS mixing angles cannot be obtained in a straightforward manner. The upshot of the discussion presented in this paper is that an accommodation to the measured neutrino properties is a non-trivial calculational exercise, because a complete scan over the parameter space of the anisotropic torsional couplings is needed.

To summarize, the quark and lepton masses have been successfully reduced to couplings among the internal spin vectors. Using these results the poor man's strategy (applied in this paper) is to choose the couplings so that the fermion masses and mixings come out right. The reader may rightfully object that everything done here is to replace one set of free parameters by another set, and one effective theory (the Standard Model) by another one (the NJL inspired internal Heisenberg model). However, as shown in ref.\cite{laa} the internal couplings can be calculated from first principles as exchange integrals over internal space just as in ordinary magnetism the exchange couplings of the Heisenberg model are in principle calculable from exchange integrals of electronic wave functions over physical space. A more ambitious program therefore is to determine all internal couplings from a fundamental theory like the higher dimensional QED considered in \cite{laa}.

\end{document}